\documentclass[aps,prl,twocolumn,superscriptaddress,preprintnumbers,showpacs]{revtex4}

\usepackage{graphics}
\begin{document}


\title{Comment on triple gauge boson interactions in \\the non-commutative electroweak sector}

\author{G.~Duplan\v{c}i\'{c}}
\affiliation{Theoretical Physics Division, 
Rudjer Bo\v{s}kovi\'{c} Institute, P.O.Box 180,
10002 Zagreb, Croatia}
\email[]{gorand@thphys.irb.hr}
\author{P.~Schupp}
\affiliation{International University Bremen, Campus Ring 1, 28759 Bremen, Germany}
\email[]{p.schupp@iu-bremen.de}
\author{J.~Trampeti\' c}
\affiliation{Theoretical Physics Division, 
Rudjer Bo\v{s}kovi\'{c} Institute, P.O.Box 180,
10002 Zagreb, Croatia}
\affiliation{Theoretische Physik, Universit\"at M\"unchen, 
Theresienstr. 37, 80333 M\"unchen, Germany}
\email[]{josipt@rex.irb.hr}

\date{\today}

\begin{abstract}
In this comment we present an analysis of electroweak 
neutral triple gauge boson couplings projected out of the  
gauge sector of the extended non-commutative standard model.
A brief overview of the current experimental situation is given.
\end{abstract}


\maketitle


The main purpose of this comment is to provide a set of electroweak 
neutral triple gauge boson (TGB) coupling constant values that are relevant
for analysis of collider physics processes.
In a previous paper \cite{behr} we presented genuine new anomalous TGB interactions,
which are not present in the standard model but
arise within the framework of the extended non-commutative standard model (NCSM) \cite{cal},
and also in the alternative approach to the NCSM given in 
\cite{Chaichian:2001py}. 
The range of coupling constant values for all TGB interactions
was not completely computed in the previous paper \cite{behr}.
Since these are necessary for any further 
collider physics analysis of the pure electroweak sector of the extended NCSM,
we are presenting a numerical analysis of all electroweak couplings in this comment.
It is observed that no two of the TGB couplings vanish simultaneously
in our model as a consequence of constraints coming from the 
values of the standard model couplings at the $M_Z$-scale. 

It is of interest for experimentalists to find TGB
couplings \cite{LEPEWWG, L3, DELPHI, OPAL, OPAL1,DELPHI1}, 
as such observations would certainly contribute to the
discovery of physics beyond the standard model (SM). In the light of the recent 
OPAL Collaboration paper \cite{OPAL1}, which presents the first limits on non-commutative 
QED obtained from collider experiments, it is obvious 
that in order to repeat the same $e^+e^- \rightarrow \gamma\gamma$ analysis 
within the framework of the extended NCSM \cite{cal,behr}, a
complete set of values of neutral triple gauge boson (TGB) coupling constants is necessary.
The non-commutative fermion-photon and fermion-Z boson couplings have been given in \cite{cal}.
Note, that strictly SM forbidden decays
coming solely from the gauge sector of the NCSM
could also be probed in high-energy collider experiments.

There are two approaches to the construction of noncommutative generalizations
of the Standard Model. The approach
\cite{Chaichian:2001py} uses some clever tricks to circumvent the problems of charge quantization and the restriction of
the noncommutative gauge group only to U(N):
It starts with an enlarged gauge group 
$U(3)\times U(2) \times U(1)$ and then removes superfluous $U(1)$ factors with the help of extra Higgs fields (higgsac's). The hypercharges and the
electric charges are still quantized but now to the correct values of the
usual quarks and leptons.
The other approach \cite{cal} to the 
NCSM solves the standard problems
of noncommutative model building with 
the help of generalized Seiberg-Witten maps. In principle the two
approaches can also be combined.
Details of the latter approach of the M\"unchen-Wien groups are given
in \cite{WESS, Zumino, asch, ws, ws1}.
We propose the use of effective Lagrangians 
constructed within the NCSM \cite{behr}, in further analysis of
scatterings of electrons and photons. It is the
only approach that allows to build models of the electroweak
sector directly based on the structure group $U(1)\times SU(2)$
in the presence of spacetime non-commutativity.

The action that we use here
should be understood as an effective theory 
up to linear order in the non-commutativity parameter $\theta$. 
New triple gauge boson (TGB) terms in the action have the following form \cite{behr}:
\begin{eqnarray}
\lefteqn{S_{gauge}=-\frac{1}{4}\int \hspace{-1mm}d^4x\, f_{\mu \nu} f^{\mu \nu}}
 \label{action2} \\
& &\hspace{-5mm}{}
-\frac{1}{2}\int \hspace{-1mm}d^4x\, {\rm Tr}\left( F_{\mu \nu} F^{\mu \nu}\right)
-\frac{1}{2}\int\hspace{-1mm} d^4x\, {\rm Tr}\left( G_{\mu \nu} G^{\mu \nu}\right)
\nonumber \\
& &\hspace{-5mm}{}
+g_s \,\theta^{\rho\tau}\hspace{-2mm}
\int\hspace{-1mm} d^4x\, {\rm Tr}
\left(\frac{1}{4} G_{\rho \tau} G_{\mu \nu} - G_{\mu \rho} G_{\nu \tau}\right)G^{\mu \nu}\nonumber \\
& &\hspace{-5mm}{}+{g'}^3\kappa_1{\theta^{\rho\tau}}\hspace{-2mm}\int \hspace{-1mm}d^4x\,
\left(\frac{1}{4}f_{\rho\tau}f_{\mu\nu}-f_{\mu\rho}f_{\nu\tau}\right)f^{\mu\nu}
 \nonumber \\
& &\hspace{-5mm}{}+g'g^2\kappa_2 \, \theta^{\rho\tau}\hspace{-2mm}\int
\hspace{-1mm} d^4x \sum_{a=1}^{3}
\left[(\frac{1}{4}f_{\rho\tau}F^a_{\mu\nu}-
f_{\mu\rho}F^a_{\nu\tau})F^{\mu\nu,a}\!+c.p.\right]
 \nonumber \\
& &\hspace{-5mm}{}+g'g^2_s\kappa_3\, \theta^{\rho\tau}\hspace{-2mm}\int
\hspace{-1mm} d^4x \sum_{b=1}^{8}
\left[(\frac{1}{4}f_{\rho\tau}G^b_{\mu\nu}-
f_{\mu\rho}G^b_{\nu\tau})G^{\mu\nu,b}\!+c.p.\right], \nonumber 
\end{eqnarray}
where $c.p.$ means cyclic permutations.
Here $f_{\mu\nu}$, $F^a_{\mu\nu}$ and $G^b_{\mu\nu}$ are the physical field strengths corresponding 
to the groups $\rm U(1)_Y$, $\rm SU(2)_L$ and $\rm SU(3)_C$, respectively. 
The constants $\kappa_1$, $\kappa_2$ and $\kappa_3$ are functions of $1/g_i^2\; (i=1,...,6)$:
\begin{eqnarray}
\kappa_1 &=& -\frac{1}{g^2_1}-\frac{1}{4g^2_2}+\frac{8}{9g^2_3}-\frac{1}{9g^2_4}+\frac{1}{36g^2_5}
+\frac{1}{4g^2_6},
\nonumber \\
\kappa_2 &=& -\frac{1}{4g^2_2}+\frac{1}{4g^2_5}+\frac{1}{4g^2_6},
\nonumber \\
\kappa_3 &=& +\frac{1}{3g^2_3}-\frac{1}{6g^2_4}+\frac{1}{6g^2_5}.
\end{eqnarray}
The $g_i$ are the coupling constants of the non-commutative electroweak sector
up to first order in $\theta$. The appearance of new coupling constants beyond
those of the standard model reflect a freedom in the strength
of the new TGB couplings.
Matching the SM action at zeroth order in $\theta$, three consistency conditions
are imposed on (\ref{action2}):
\begin{eqnarray}
\frac{1}{{g'}^2} &=& \frac{2}{g^2_1}+\frac{1}{g^2_2}+\frac{8}{3g^2_3}+\frac{2}{3g^2_4}+\frac{1}{3g^2_5}
+\frac{1}{g^2_6},
\nonumber \\
\frac{1}{g^2}&=& \frac{1}{g^2_2}+\frac{3}{g^2_5}+\frac{1}{g^2_6},\nonumber \\
\frac{1}{g_s^2}&=& \frac{1}{g^2_3}+\frac{1}{g^2_4}+\frac{2}{g^2_5}.
\label{L2}
\end{eqnarray}
>From the action (\ref{action2}) we extract the genuine new 
neutral triple-gauge boson terms which are not present in the SM Lagrangian. 
In terms of the SM physical fields $G$, $A$ and $Z$, they are \cite{behr}
\begin{eqnarray}
{\cal L}_{\gamma\gamma\gamma}&=&\frac{e}{4} \sin2{\theta_W}\;{\rm K}_{\gamma\gamma\gamma}
{\theta^{\rho\tau}}A^{\mu\nu}\left(A_{\mu\nu}A_{\rho\tau}-4A_{\mu\rho}A_{\nu\tau}\right),
\nonumber\\
{\rm K}_{\gamma\gamma\gamma}&=&\frac{1}{2}\; gg'(\kappa_1 + 3 \kappa_2);  
\nonumber\\
{\cal L}_{Z\gamma\gamma}&=&\frac{e}{4} \sin2{\theta_W}\,{\rm K}_{Z\gamma \gamma}\,
{\theta^{\rho\tau}}
\left[2Z^{\mu\nu}\left(2A_{\mu\rho}A_{\nu\tau}-A_{\mu\nu}A_{\rho\tau}\right)\right.\nonumber\\
& & +\left. 8 Z_{\mu\rho}A^{\mu\nu}A_{\nu\tau} - Z_{\rho\tau}A_{\mu\nu}A^{\mu\nu}\right], 
\nonumber \\
{\rm K}_{Z\gamma\gamma}&=&\frac{1}{2}\; \left[{g'}^2\kappa_1 + \left({g'}^2-2g^2\right)\kappa_2\right]; 
\nonumber\\
{\cal L}_{ZZ\gamma}&=&{\cal L}_{Z\gamma\gamma}(A\leftrightarrow Z),
\nonumber \\
{\rm K}_{ZZ\gamma}&=&\frac{-1}{2gg'}\; \left[{g'}^4\kappa_1 + g^2\left(g^2-2{g'}^2\right)\kappa_2\right]; 
\nonumber\\ 
{\cal L}_{ZZZ}&=&{\cal L}_{\gamma\gamma\gamma}(A\to Z),
\nonumber\\
{\rm K}_{ZZZ}&=&\frac{-1}{2g^2}\; \left[{g'}^4\kappa_1 + 3g^4\kappa_2\right]; 
\nonumber\\
{\cal L}_{Zgg}&=&{\cal L}_{Z\gamma\gamma}(A\to G^b), 
\nonumber \\
{\rm K}_{Zgg}&=&\frac{g^2_s}{2} \left[1+(\frac{{g'}}{g})^2\right]\kappa_3; 
\nonumber\\
{\cal L}_{\gamma gg}&=&{\cal L}_{Zgg}(Z\rightarrow A),
\nonumber \\
{\rm K}_{\gamma gg}&=&\frac{-g^2_s}{2}\;
\left[\frac{g}{g'}+\frac{g'}{g}\right]\kappa_3,
\label{L6}
\end{eqnarray} 
where $A_{\mu\nu} \equiv \partial_{\mu}A_{\nu} -
\partial_{\nu}A_{\mu}$, etc.

\begin{figure}
 \resizebox{0.45\textwidth}{!}{%
  \includegraphics{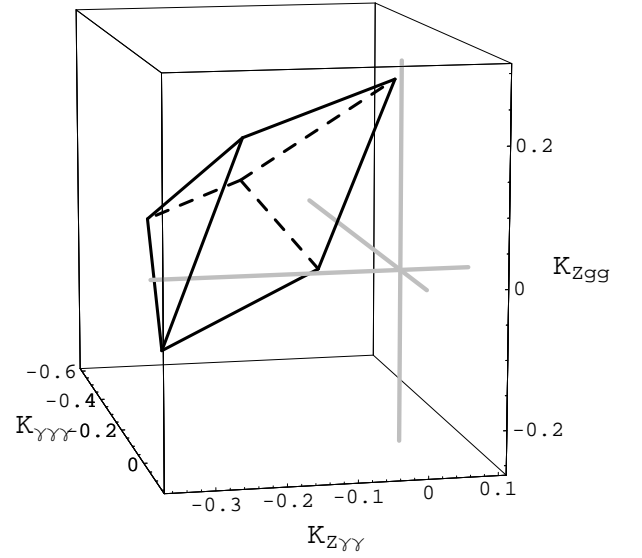}}
 \caption{The three-dimensional pentahedron that bounds possible values
 for the coupling constants ${\rm K}_{\gamma\gamma\gamma}$, 
 ${\rm K}_{Z\gamma\gamma}$ 
 and ${\rm K}_{Zgg}$ at the $M_Z$ scale. 
 }
 \label{fig1}
\end{figure}

The above three conditions (\ref{L2}), together with the requirement that 
$1/g_i^2 > 0$, define a three-dimensional pentahedron in
the six-dimensional moduli space spanned by $1/g_1^2,...,1/g_6^2$. 
It is possible to express all constraints and conditions in terms of 
coupling constants ${\rm K}_{\gamma\gamma\gamma}$, etc. 
Fig.\ref{fig1} shows the three-dimensional pentahedron \cite{behr} that bounds 
the allowed values for the dimensionless coupling constants 
${\rm K}_{\gamma\gamma\gamma}$, ${\rm K}_{Z\gamma\gamma}$ 
and ${\rm K}_{Zgg}$, at the scale $M_Z$.
For any chosen point within the pentahedron 
in Fig.\ref{fig1}, the remaining three coupling constants 
${\rm K}_{Z Z \gamma}$, ${\rm K}_{Z Z Z}$ 
and ${\rm K}_{\gamma g g}$, are uniquely fixed by the equations
\begin{eqnarray}
{\rm K}_{ZZ\gamma} &=&
\frac{1}{2} \left(\frac{g}{g'}-3\frac{g'}{g}\right){\rm K}_{Z\gamma\gamma} 
-\frac{1}{2}\left(1-\frac{g^{\prime 2}}{g^2}\right){\rm K}_{\gamma\gamma\gamma},
\nonumber \\
{\rm K}_{ZZZ} &=&
\frac{3}{2} \!\left(1-\frac{g^{\prime 2}}{g^2}\right){\rm K}_{Z\gamma\gamma} 
-\frac{1}{2}\frac{g'^2}{g^2}\!\left(3-\frac{g^{\prime 2}}{g^2}\right){\rm K}_{\gamma\gamma\gamma},
\nonumber \\
{\rm K}_{\gamma gg} &=& -\frac{g}{g'}{\rm K}_{Z gg}.
\label{L7}
\end{eqnarray}
The values for all six coupling constants at the pentahedron vertices are given in Table I.
>From Table I it is possible to construct the allowed region for any pair of couplings:
${\rm K}_{\gamma\gamma\gamma}$, ${\rm K}_{Z\gamma\gamma}$, ${\rm K}_{ZZ\gamma}$, 
${\rm K}_{ZZZ}$, ${\rm K}_{\gamma gg}$ and ${\rm K}_{Z\gamma\gamma}$.
\renewcommand{\arraystretch}{2.5}
\begin{table}
\caption{The values of the triple gauge boson couplings at the vertices of the
pentahedron in the extended NCSM at the $M_Z$ scale.}
\vspace{0.5cm}
\begin{tabular}{|c|c|c|c|c|c|}
\hline
${\rm K}_{\gamma\gamma\gamma}$ & ${\rm K}_{Z\gamma\gamma}$ & ${\rm K}_{Zgg}$ & ${\rm K}_{ZZ\gamma}$ & ${\rm K}_{ZZZ}$ & ${\rm K}_{\gamma gg}$\\
\hline \hline
$ -0.184 $ & $ -0.333 $ & $ 0.054 $ & $ 0.035 $ & $ -0.213 $ & $ -0.098 $\\
\hline
$ -0.027 $ & $ -0.340 $ & $ -0.108 $ & $ -0.021 $ & $ -0.337 $ & $ 0.197 $\\
\hline
$ 0.129 $ & $ -0.254 $ & $ 0.217 $ & $ -0.068 $ & $ -0.362 $ & $ -0.396 $\\
\hline
$ -0.576 $ & $ 0.010 $ & $ -0.108 $ & $ 0.202 $ & $ 0.437 $ & $ 0.197 $\\
\hline
$ -0.497 $ & $ -0.133 $ & $ 0.054 $ & $ 0.162 $ & $ 0.228 $ & $ -0.098 $\\
\hline
$ -0.419 $ & $ 0.095 $ & $ 0.217 $ & $ 0.155 $ & $ 0.410 $ & $ -0.396 $\\
\hline
\end{tabular}
\label{t:tab1}
\end{table}
\renewcommand{\arraystretch}{1}
The range of values for a full set of electroweak coupling constants is given in Figs.~2 to 7. 
\begin{figure}
 \resizebox{0.45\textwidth}{!}{%
  \includegraphics{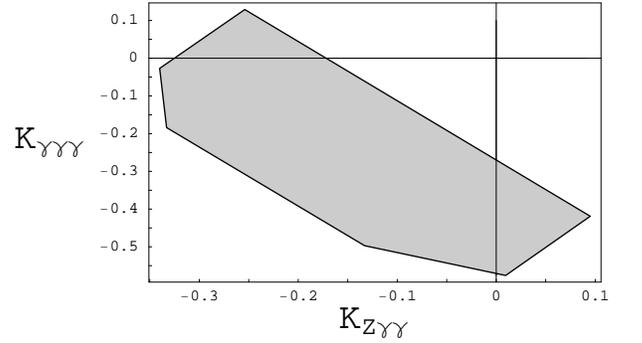}}
 \caption{The allowed region for (${\rm K}_{Z\gamma\gamma},\,{\rm K}_{\gamma\gamma\gamma}$)  
 couplings. 
 }
\end{figure}

\begin{figure}
 \resizebox{0.45\textwidth}{!}{%
  \includegraphics{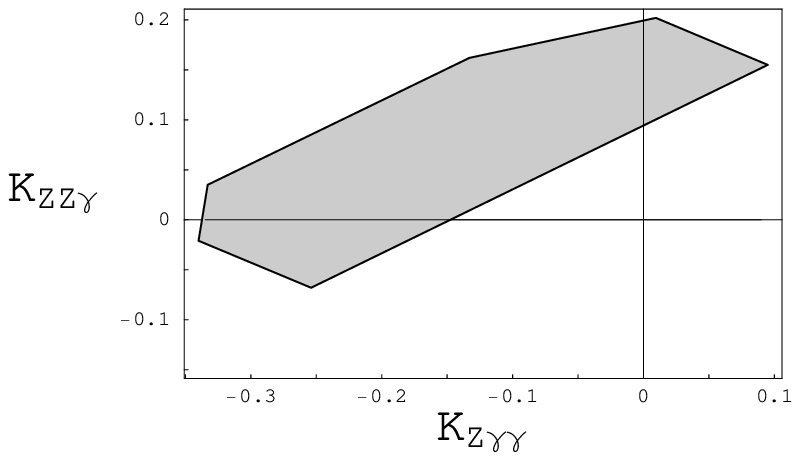}}
 \caption{The allowed region for (${\rm K}_{Z\gamma\gamma},\,{\rm K}_{ZZ\gamma}$). 
 }
\end{figure}

\begin{figure}
 \resizebox{0.45\textwidth}{!}{%
  \includegraphics{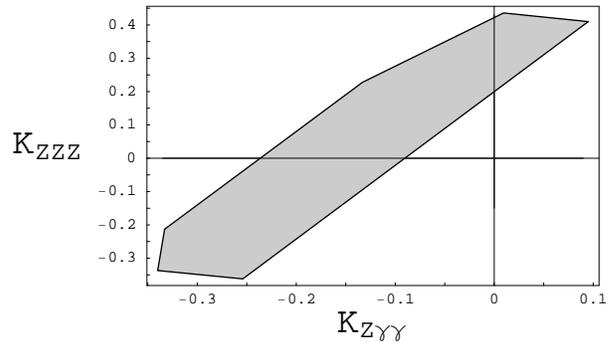}}
 \caption{The allowed region for (${\rm K}_{Z\gamma\gamma},\,{\rm K}_{ZZZ}$). 
 }
\end{figure}

\begin{figure}
 \resizebox{0.45\textwidth}{!}{%
  \includegraphics{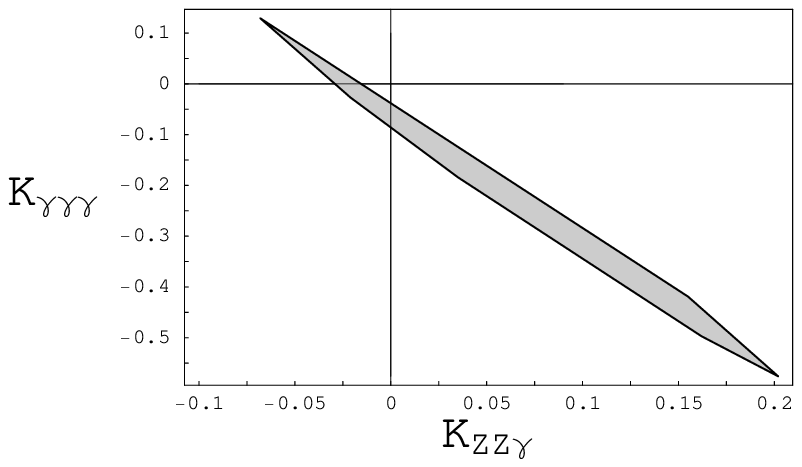}}
 \caption{The allowed region for (${\rm K}_{ZZ\gamma},\,{\rm K}_{\gamma\gamma\gamma}$). 
 }
\end{figure}

\begin{figure}
 \resizebox{0.45\textwidth}{!}{%
  \includegraphics{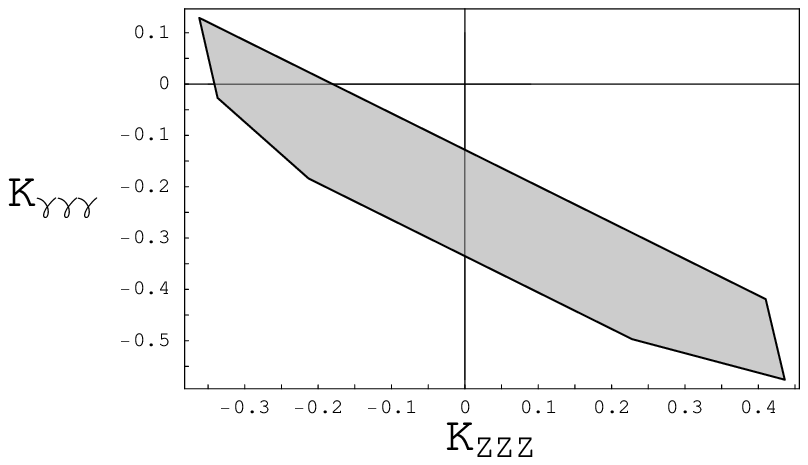}}
 \caption{The allowed region for (${\rm K}_{ZZZ},\,{\rm K}_{\gamma\gamma\gamma}$). 
 }
\end{figure}

\begin{figure}
 \resizebox{0.45\textwidth}{!}{%
  \includegraphics{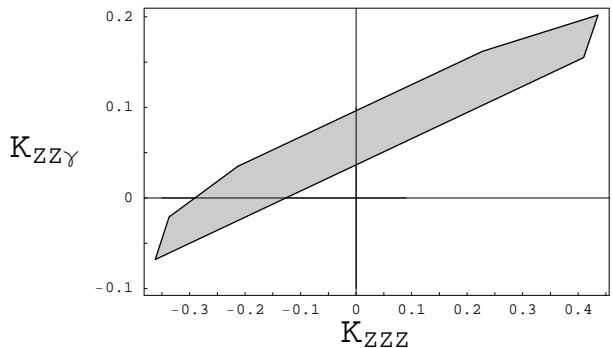}}
 \caption{The allowed region for (${\rm K}_{ZZZ},\,{\rm K}_{ZZ\gamma}$). 
 }
\end{figure}
The important property evident from Eq. (\ref{L7}) and
Figs.~2 to 7 is that any combination of two TGB
coupling constants from the gauge sector can never vanish 
simultaneously due to the constraint set by the value of the SM coupling
constants at the $\rm{M_Z}$ scale. 



Evidence for non-commutativity coming from the gauge sector
should be searched for in processes involving the above couplings. 
The experimental discovery of the kinematically allowed $Z\rightarrow \gamma\gamma$ decay
would indicate a violation of the Yang theorem and would be a
possible signal of space-time non-commutativity. This would fix 
the quantity $|{\rm K}_{Z\gamma\gamma}/\Lambda^2_{\rm NC}|^2$, where $\Lambda_{\rm NC}$
represents the scale of non-commutativity.
Inclusion of other triple-gauge boson interactions through $2\to 2$ scattering experiments would 
sufficiently reduce the available parameter space of our model. 


To get an idea about the order-of-magnitude of the rate, 
let us choose the central value of the $Z\gamma\gamma$
coupling constants ${|{\rm K}_{Z\gamma \gamma}|\simeq 1/10}$ and assume that maximal non-commutativity
occurs at the scale of $\sim$ 1 TeV.
The resulting branching ratio \cite{behr} for this decay, at tree level, would 
then be $BR(Z\to \gamma\gamma)\simeq 4\times 10^{-8}$.

Concerning the question of the scale of non-commutativity $\Lambda_{NC}$ in forbidden decays
and in scatterings of electrons and photons, the experimental situation can be summarized as follows:

(i) The joint efforts of the DELPHI, ALEPH, OPAL and L3 Collaborations 
\cite{LEPEWWG, L3, OPAL, OPAL1} give us hope that in 
not too much time all collected data from the LEP experiments will be counted and analysed, producing 
tighter bounds on triple-gauge boson couplings and the scale of non-commutativity, i.e. on the 
quantities like $|{\rm K}_{Z\gamma\gamma}/\Lambda^2_{NC}|^2$, etc.

(ii) The sensitivity to the NC parameter $\theta^{\mu\nu}$ could be in the range of the next
generation of linear colliders, with a c.m.e. around a few TeV. 

(iii) The first limits on the NCQED obtained from collider experiments were recently 
presented by the OPAL Collaboration in \cite{OPAL1}. They found no significant deviation from 
the SM prediction and at the 95\% confidence level the limit was set on the non-commutative scale
$\Lambda_{NC}\;>$  141 GeV. This is valid for all relevant angles that determine a unique direction in space.

(iv) Finally, note that the best testing ground for studies of anomalous TGB couplings, 
before the start of the linear $e^+e^-$ collider, will be the LHC \cite{muller}.

In conclusion, the gauge sector of the SU(2)$\times$U(1) is a possible place for the experimental discovery
of space-time non-commutativity.
We believe that the importance of a possible
discovery of space-time non-commutativity at very short distances would convince
collider particle physics experimentalists to search further for the forbidden decay $Z\to \gamma\gamma$, as well as
for other anomalous triple neutral gauge boson couplings.
\\
\begin{acknowledgments}
We would like to thank Th. M\" uller, V. Ruhlmann-Kleider for helpful discussions.
This work was supported by the Ministry of Science and Technology of 
the Republic of Croatia under Contract No. 0098002.
\end{acknowledgments}

\end{document}